\documentclass[A4,10pt]{article} 

\usepackage{osameet3} 

\usepackage{amsmath,amssymb}
\usepackage[colorlinks=true,bookmarks=false,citecolor=blue,urlcolor=blue]{hyperref} 

\usepackage{subcaption}

\usepackage{mathtools, bm}
\usepackage{amssymb, bm}
\usepackage{framed}
\usepackage{textcomp}
\usepackage{stmaryrd}

\usepackage{xspace}

\newcommand{\ket}[1]{\ensuremath{|#1\rangle}\xspace}

\begin{document}

\title{Technology Trends for Mixed QKD/WDM Transmission up to 80 km}

\author{Romain All\'eaume, Rapha\"el Aymeric, C\'edric Ware, Yves Jaou\"en}
\address{T\'el\'ecom Paris, Institut Polytechnique de Paris, 46 Rue Barrault, 75013 Paris}
\email{romain.alleaume@telecom-paris.fr}

\copyrightyear{2020}

\begin{abstract}
 We give a survey of some of the recent progress made in deploying quantum and classical communications over a shared fiber, focusing in particular on results obtained using continuous-variable QKD. \end{abstract}

\vspace{0.4 cm}

\paragraph{Introduction} Quantum Key Distribution (QKD) is the only technology allowing to
distribute a  key over a public channel, with security  even against a computationally unlimited attacker. It consists in encoding classical information into quantum signals that are exchanged between two parties connected by a quantum channel, supplemented by a classical authenticated channel. The crucial benefit of quantum signals is that they are sensitive to eavesdropping. This allows to design QKD protocols that distill secret keys, whose amount is
limited by the noise.  Monitoring the noise gives a bound on the rate of key generation guaranteed to be eavesdropping-free.

Most of the effort on QKD system design and most of the experimental demonstrations have  been realized on dark fiber. This however restricts the deployability of QKD to a limited number of scenarios where the barriers of availability and price for reserving a dedicated dark fiber for QKD could both be overcome. Wavelength Division Multiplexing (WDM) compatibility of QKD would thus imply a significant improvement for QKD in terms of cost-effectiveness and compatibility with existing optical infrastructures.
However, WDM coexistence of QKD with intense classical channels raises some new challenges because of the additional noise induced onto the quantum channel. This noise can be due to insufficient isolation between the classical and quantum channels, which can be managed by more severe isolation. On the other hand, non-linear effects and in particular spontaneous Raman scattering leads to wide-spectrum (over 200 nm) of spurious noise photons, some of which being inevitably spectrally matched with the quantum channel.
Coping with Raman noise induced by classical channels is a major challenge for QKD systems and especially for discrete variable QKD (DV-QKD) that rely on single photon detectors that are not spectrally selective: Raman scattering spectrum leads to in-band noise photons that cannot be efficiently removed by wavelength filters without adding significant extra losses.

 \paragraph{Early work }
 Pioneering work on QKD in WDM environment has been performed with DV-QKD systems, in coarse-WDM \cite{Townsend:elelett97} as well as later in dense-WDM configurations \cite{peters:njp09, chapuran:njp09}, however over distances below 25 km.
Several DV-QKD experiments have then  tried to extend the distance for mixed QKD/WDM. In \cite{eraerds:njp12}, 4 classical channels
 where multiplexed with a DV-QKD system and 50 km operation was demonstrated. However,
 the input power of the classical channels was attenuated to the smallest
  possible power compatible with the sensitivity limit of the optical receiver (-26 dBm).
  Attenuating the classical channel launch power was also used in  \cite{patel:prx12}  where the impressive distance of 90 km was demonstrated, however with an input power limited down to -18.5 dBm and in addition the use of temporal filtering techniques. Temporal and spectral filtering techniques have also been optimized in \cite{patel:apl14} to allow the first demonstration of DV-QKD in coexistence with one 0 dBm classical channel, at 25 km.
    
      \paragraph{Continuous Variable QKD (CV-QKD)}   
  
As analyzed in \cite{qi:njp10}, the coherent detection used in CV-QKD to measure the field quadratures acts as a natural and extremely selective filter whose acceptance is equal to the bandwidth of the coherent detector, i.e. typically ranging  between 1 and a few hundreds of MHz. As a consequence, CV-QKD can  operate in a regime where it filters out spurious light down to a single spatio-temporal mode. This feature allows us to achieve results that could not have been obtained so far with DV-QKD, namely the coexistence of a fully operational CV-QKD system over metropolitan distance with an intense dense-wavelength-mutiplexed classical channel of several dBm.

\begin{figure}[htb!]
        \begin{subfigure}[b]{0.54\textwidth}
                \includegraphics[width=\textwidth]{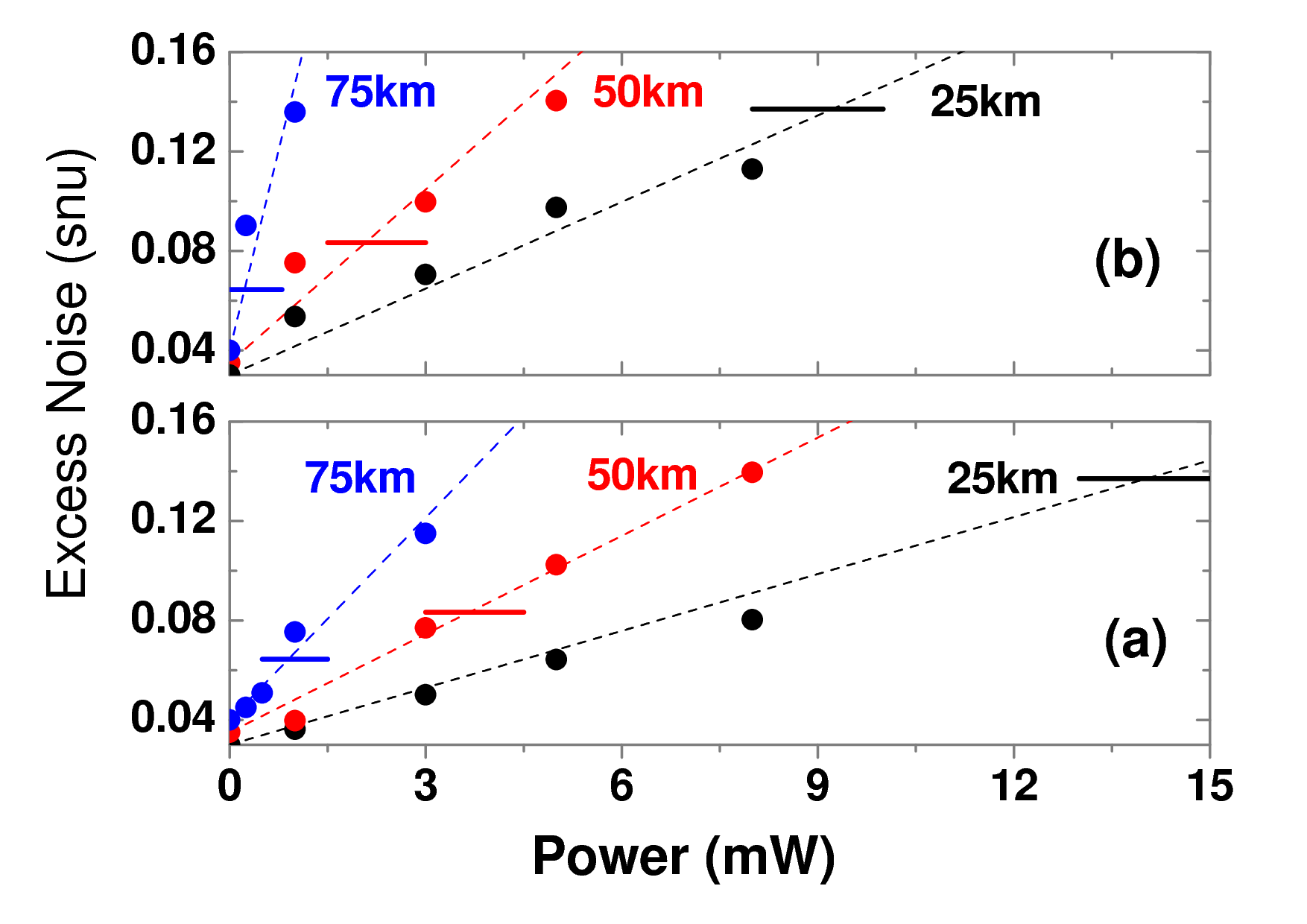}
               \label{figure:Result}
        \end{subfigure}
                \begin{subfigure}[b]{0.48\textwidth}
                \includegraphics[width=\textwidth]{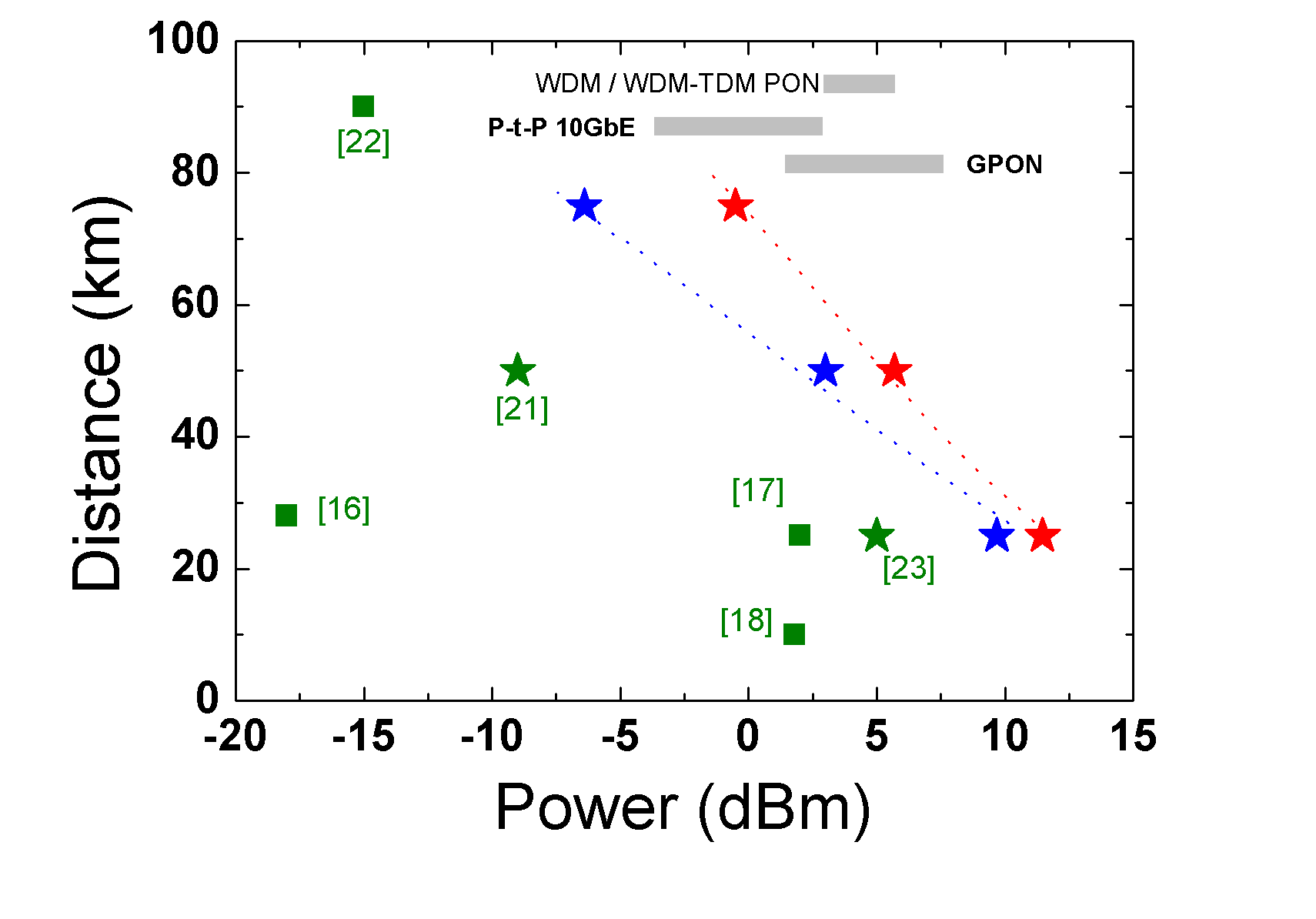}
                \label{figure:CompareWDMworks}
        \end{subfigure}
            \vspace{-9mm}
        \caption{\textbf{(Left)}: \textbf{Excess noise measurements vs launch power in forward (a)  and backward  (b) channel configuration}.
  Black, red and blue data points are the  excess noise  evaluated at Alice
  for fiber length of 25km, 50km and 75km, for different classical channel power.
    Dashed lines indicate the expected excess noise curve and solid horizontal
  lines are null key threshold for CV-QKD, for the respective channel distance.\\
          \textbf{(Right): Tolerable classical channel power vs Reachable distance}: Performance of QKD in the context
  of coexistence with classical optical channels.
  Red and blue colors represents our results with a CV-QKD system, in forward and backward classical channel configuration,
  while previous
  works with DV-QKD systems are in Green.
  Stars: experiments conducted in the C-band (DWDM). Squares: experiments conducted in CWDM. The dotted red and blue lines are the forward and backward simulation curve for the null key rate in
  the current experiment. Gray bands show transmitter input power range in different standardized optical networks. Figure taken from \cite{KQA15}, see original article for the number-reference correspondance.}
                \label{doublefigure}
    \vspace{-6mm}
\end{figure}

In \cite{KQA15} we have experimentally validated the capacity of CV-QKD to co-propagate with intense WDM signals.
 Our experimental test-bed consisted in a CV-QKD link (operated at 1531.12 nm) multiplexed
with one DWDM classical channel whose wavelength is set at 1550.12nm. We could check, as displayed on Fig.\ref{doublefigure}/left, the linear dependence of Raman-induced noise with launch power, and test CV-QKD operation at 25, 50 and 75 km.  We also observe for example that up to 14 mW (11.5 dBm) of launch power can be tolerated by CV-QKD in the forward configuration, at 25 km.

CV-QKD can be deployed in coexistence with classical channels of unprecedented power levels thanks to the
mode selection property of its coherent detection. This gives CV-QKD
an advantage for the integration into different optical network architectures and in particular access networks. 
Figure \ref{doublefigure}/right displays a comparison between
DV-QKD and CV-QKD in terms of tolerable classical channel power.
As it can be seen, CV-QKD can be integrated into different high
power passive optical networks such as for example Gigabit PON, 10G-PON and WDM/TDM PON. On
the other hand, integration of DV-QKD in  such optical networks
requires either some modifications in the architecture or further advances in the noise
reduction techniques applicable to DV-QKD.\\

    \vspace{-3mm}

 \paragraph{Towards QKD/WDM coexistence over optical backbone links }   

The ability to deploy QKD over optical backbones or over inter-datacenter links could be a game-changer for the development of the technology, significantly reducing QKD deployment cost overhead  and most importantly opening radically larger and security-relevant market segments.
This objective however comes with very stringent requirements: \\1)  Cover a distance equivalent to the typical optical fiber span, i.e. around 80 km ; \\2) QKD operation with positive key rates in coexistence with several WDM classical channels (up to 100 in backbone) each of nominal (0 dBm) launch power, hence requiring ultra low WDM-induced noise ;\\ 3)  QKD integration should ideally have a minimal impact over standard WDM link information transmission capacity. \\

\begin{figure}[h]
\centering
 \includegraphics[width=140mm]{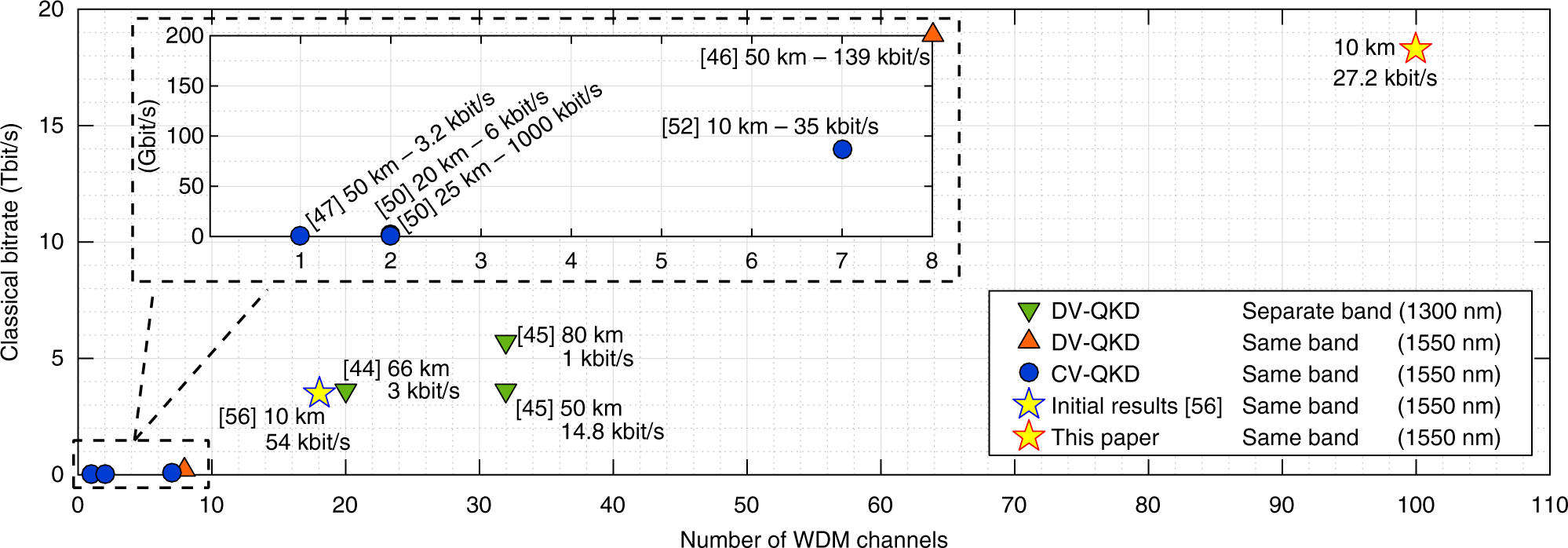}
 \vspace{-1mm}
  \caption{Figure taken from \cite{Eriksson2019} and comparing the total classical bitrate, the number of wavelength division multiplexing channels  and the total data-rate of classical channels. It illustrates the abililty to operate CV-QKD in mixed WDM environment close to the backbone regime, i.e with 100 classical channels and Terabit/s classical capacity. However the demonstrated distance was only 10 km, and the launch power of each classical channel is approx. -7 dBm. See original article for details, and the number-reference correspondance.}
   \label{figure:erikson}
    \vspace{-4mm}
\end{figure}

Although these requirements have not yet been fulfilled in a single experimental demonstration, significant steps have been recently made. 
In \cite{Eriksson2019}, CV-QKD co-propagation jointly transmitted with 100 WDM channels over which a datarate of 18.3 Tbit/s was being sent, over a realistic set-up. This impressive demonstration  meets criteria 2) and 3), however was demonstrated only over 10 km. Fig. \ref{figure:erikson} compares this figure with earlier demonstrations. At OFC 2019, Kleis et. al. \cite{Kleis2019} reported on an experimental demonstration of mixed CV-QKD /WDM with classical signals placed in the S-band. This has allowed them to multiplex up to 28 classical channels at 0 dBm each (approx. 14 dBm of total power) in coexistence with CV-QKD, i.e a notable progress towards criteria 1) and 2). 
 Finally, we notice that \cite{Mao2018} has made decisive steps in meeting criteria 1-2-3) all along, by demonstrating the integration of DV-QKD, in coexistence with 21 dBm of classical signals, carrying 3.6 Tbs data-rate, over 66 km. This record performance however relies on the use of large-core fibers and cannot therefore be directly applied within existing networks.

\paragraph{Conclusion and Perspectives}

The performance boundary of mixed QKD/WDM has sharply increased over the past years, driven notably by the progress of QKD systems performance and maturity. WDM coexistence can be envisaged over fiber links as long as 100 km \cite{Dynes16} using DV-QKD (however with a classical launch power reduced to -25.5 dBm), and in coexistence with several tens \cite{KQA15, Kleis2019}  of standard 0 dBm channels, using CV-QKD (however over shorter distances, i.e in the  25-50 km range). It remains an interesting scientific and engineering challenge to see how the strength of DV and CV quantum communications systems can be optimally combined to push the  performance boundary further, and in particular up to the tipping point where QKD systems could be deployed over existing backbone links.

\paragraph{Acknowledgements}
This project has received funding from the European Union's Horizon 2020 research and innovation programme under grant agreement CIVIQ No 820466.


\begin{thebibliography}{99} 

\bibitem{Townsend:elelett97} P.D.Townsend, "Simultaneous quantum cryptographic key distribution and conventional data transmission over installed fibre using wavelength-division multiplexing." \textit{Electronics Letters} 33, 188 (1997).
\bibitem{peters:njp09} N. A. Peters, \emph{et~al.}, Peters, N. A., et al. "Dense wavelength multiplexing of 1550 nm QKD with strong classical channels in reconfigurable networking environments." \textit{New Journal of physics} 11.4, 045012 (2009).
\bibitem{chapuran:njp09} T.E. Chapuran,  \emph{et~al.},  "Optical networking for quantum key distribution and quantum communications." \textit{New Journal of Physics} 11, 105001 (2009).
\bibitem{eraerds:njp12} P. Eraerds \emph{et~al.},  "Quantum key distribution and 1 Gbps data encryption over a single fibre." \textit{New Journal of Physics} 12,  063027 (2010).
\bibitem{patel:prx12} K.A. Patel \emph{et~al.}, "Coexistence of high-bit-rate quantum key distribution and data on optical fiber." \textit{Physical Review X} 2, 041010 (2012).
\bibitem{patel:apl14}K.A. Patel \emph{et~al.}, "Quantum key distribution for 10 Gb/s dense wavelength division multiplexing networks." \textit{Applied Physics Letters} 104, 051123 (2014).
\bibitem{qi:njp10} B. Qi,  \emph{et~al.}  "Feasibility of quantum key distribution through a dense wavelength division multiplexing network." \textit{New Journal of Physics} 12, 103042 (2010).

\bibitem{KQA15} R. Kumar, \emph{et~al.},  "Coexistence of continuous variable QKD with intense DWDM classical channels." \textit{New Journal of Physics} 17, 043027 (2015). 

\bibitem{Eriksson2019} Eriksson, Tobias A., \emph{et~al.},  "Wavelength division multiplexing of continuous variable quantum key distribution and 18.3 Tbit/s data channels." \textit{Communications Physics} 2, 9 (2019).


 \bibitem{Kleis2019} S. Kleis \emph{et~al.},  "Experimental Investigation of Heterodyne Quantum Key Distribution in the S-Band Embedded in a Commercial DWDM System." \textit{Optical Fiber Communication Conference}, OSA, (2019).

\bibitem{Mao2018}  Mao, Yingqiu, \emph{et~al.},. "Integrating quantum key distribution with classical communications in backbone fiber network." \textit{Optics express} 26, 6010-6020 (2018).







\bibitem{Dynes16} J.F. Dynes  \emph{et~al.}, "Ultra-high bandwidth quantum secured data transmission." Scientific reports 6, 35149 (2016).


\end{thebibliography}
\end{document}